\documentclass[11pt]{article}
\usepackage{amsmath,amsthm,amscd,array,amssymb}

\begin{document}

\begin{center}
{\Large{\bf Topological gauge theories
\\
\medskip\smallskip
with antisymmetric tensor matter fields}}
\\
\bigskip\medskip
{\large{\sc B. Geyer}}$^a$
\footnote{Email: geyer@itp.uni-leipzig.de}
and 
{\large{\sc D. M\"ulsch}}$^{b}$
\footnote{Email: muelsch@informatik.uni-leipzig.de}
\\
\smallskip
{\it $^a$ Universit\"at Leipzig, Naturwissenschaftlich-Theoretisches Zentrum
\\
$~$ and Institut f\"ur Theoretische Physik, D--04109 Leipzig, Germany
\\\smallskip
$\!\!\!\!\!^b$ Wissenschaftszentrum Leipzig e.V., D--04103 Leipzig, Germany}
\\
\end{center}
\begin{abstract}
\noindent
{\small {A new type of topological matter interactions involving
second--rank antisymmetric tensor matter fields with an underlying
$N_T \geqq 1$ topological supersymmetry are proposed. 
The construction of the 4--dimensional, $N_T = 1$ Donaldson--Witten theory, 
the $N_T = 1$ super--BF model and the $N_T = 2$ topological B--model 
with tensor matter are explicitly worked out.}}
\end{abstract}


\section{Introduction}

It has been known for a long time that in quantum field and string theories
besides totally symmetric tensor and tensor--spinor fields also second--rank
antisymmetric tensor fields play an important role. As a significant example,
the Green--Schwarz anomaly cancellation mechanism \cite{1} underlies, 
among others, a coupling of a two--form gauge potential with a Chern--Simons 
form. 

All the known couplings in Minkowski space--time involving 
antisymmetric tensor fields may be put into two categories. Depending on 
whether they transform as gauge or as matter fields, one distinguishes 
\\
(i) tensor gauge couplings,
\begin{align}
\label{1.1}
S_{\rm gauge} &\propto \int_M d^4x\, ( \partial^a \tilde{B}_{ac} ) 
( \partial_b \tilde{B}^{bc} ),
\qquad
\tilde{B}_{ab} = \frac{i}{2} \epsilon_{abcd} B^{cd},
\qquad
\Tilde{\Tilde{B}}_{ab} = - B_{ab},
\\
\intertext{(ii) (conform invariant) tensor matter couplings,}
\label{1.2}
S_{\rm matter} &\propto \int_M d^4x\, ( \partial^a \varphi_{ac} ) 
( \partial_b \varphi^{bc} )^\dagger,
\qquad
\varphi_{ab} = T_{ab} + i \tilde{T}_{ab},
\end{align} 
where $\varphi_{ab}$ is an antisymmetric complex tensor field 
involving the tensor matter field $T_{ab}$ and satisfying the complex 
self--duality condition $\varphi_{ab} = i \tilde{\varphi}_{ab}$. 

In the first case, the action (\ref{1.1}) possesses a first--stage reducible 
gauge symmetry $\delta_G B_{ab} = \partial_{[a} \omega_{b]}$ \cite{2}. Such
antisymmetric tensor gauge fields appear quite naturally in extended 
supergravity theories \cite{3} and in effective low--energy tensor gauge 
theories derived from string models \cite{4}, e.g., the axion/dilaton
complex in Calabi--Yau compactifications of type--II superstrings \cite{5}. 

In the second case, gauge symmetry is lost, but the action (\ref{1.2}) 
exhibits an invariance under the (global) chiral symmetry 
$\delta_C T_{ab} = \alpha \tilde{T}_{ab}$, i.e., $T_{ab}$ transforms as 
an ordinary matter field. Antisymmetric tensor matter fields arise in 
extended conformal supergravity theories \cite{6} and in 2D conformal quantum 
field theories (CQFT's) \cite{7}.

Besides of this, there is a renewed interest in antisymmetric tensor 
fields due to their connection to a large class of Schwarz type topological 
models, namely the BF--models \cite{8}, which are exactly solvable QFT's. 
Generally, topological quantum field theories (TQFT's) \cite{9} are 
characterized by observables depending only on the global features of the 
manifold on which they are defined and provide novel representations of 
certain topological invariants. The most familar examples, which have 
been widely studied during the last years, are the Donaldson--\break Witten 
theory \cite{10}, the Chern--Simons gauge theory \cite{11} in $D = 3$ and the 
topological sigma models \cite{12} in $D = 2$, which constitute quantum field 
theoretic representations of the theory of Donaldson invariants \cite{13},
of knot and link invariants \cite{14} and of Gromov invariants \cite{15}, 
respectively. Moreover, TQFT's have significantly enhanced our understanding 
of CQFT's in $D = 2$ and they promised new insights into string 
theories \cite{16}.  

In the Schwarz type topological models the antisymmetric tensor fields appear 
always as gauge fields. The aim of the present paper is to construct, rather 
differently, Witten type topological models which include antisymmetric 
tensors fields as matter fields. More precisely, we consider extensions of 
the 4D, $N_T = 1$ Donaldson--Witten (DW) theory, the $N_T = 1$ 
super--BF model \cite{17} and the $N_T = 2$ topological B--model, 
constructed by Marcus \cite{18}, respectively, which involve a coupling of 
the gauge field $A_\mu$ to the (anti)self--dual parts $T_{\mu\nu}^\pm$ of a 
second--rank antisymmetric matter field $T_{\mu\nu}$. These models allow, 
in principle, also the inclusion of a quartic tensor self--interaction term. 
Both types of interaction terms, when put on a general curved 4--dimensional 
gravitational background with Euclidean signature, may be regarded as a 
non--abelian generalization and $N_T \geqq 1$ supersymmetric extension of the 
abelian axial gauge model for antisymmetric tensor matter fields in Minkowski 
space--time introduced by Avdeev and Chizhov \cite{19}. 

The search for a new type of topological matter action is motivated
as follows: 

One of the possible constructions of DW theory consists in twisting the 
action of Euclidean $N = 2$ super Yang--Mills (SYM) theory with 
global automorphism group $Sp(2) \otimes U(1)$ ($R$--symmetry) and 
Euclidean rotation group $SO(4) \cong SU(2)_L \otimes SU(2)_R$ by replacing 
$SU(2)_L$ through the diagonal subgroup of $SU(2)_L \otimes Sp(2)$ 
and coupling the theory to Euclidean gravity \cite{10}. 
Due to its independence on the gauge coupling constant $e$ there exists the 
possibility to study the observables of the theory from both
the perturbative and non--perturbative point of view, i.e., either in the 
weak or in the strong coupling limit, $e \rightarrow 0$ or 
$e \rightarrow \infty$, respectively. Perhaps the most important outcome
of both approaches is the existence of a totally unexpected relation between 
two different moduli spaces in 4D topology, one defined by the anti--selfdual
instanton equations \cite{20} and another one defined by the abelian
Seiberg--Witten monopol equations \cite{21}. These moduli problems can be
naturally generalized including also spinor fields, namely by twisting the 
$N = 2$ SYM coupled to $N = 2$ {\it matter hypermultiplets} in various 
representations of the gauge group \cite{22}. Since the global 
$R$--symmetry group of $N = 2$ supersymmetric gauge theories 
is at most $U(2) \cong Sp(2) \otimes U(1)$, the twist in these more general 
cases is essentially unique. The moduli space associated to that 
generalized DW theory is determined by the non--abelian monopol equations.

To construct fundamentally different, $N_T > 1$ topological theories, 
one needs at least $N = 4$ Euclidean SYM. Since the $R$--symmetry
group of $N = 4$ supersymmetric gauge theories is $SU(4)$ there exist three 
non--equivalent ways of twisting the Euclidean rotation group with 
the $R$--symmetry group. One of them, the $N_T = 2$ topological A--model,
constructed by Yamron \cite{23}, was studied by Vafa and Witten in order
to perform a strong coupling test of S--duality \cite{24}. Another one, 
the $N_T = 2$ topological B--model leads to a theory 
whose moduli space is dominated by flat complexified gauge fields 
$A_\mu \pm i V_\mu$; it can be regarded as a deformation of the $N_T = 1$ 
super--BF model \cite{25}. The remaining one, also constructed by Yamron 
\cite{23}, is the $N_T = 1$ half--twisted theory which provides another 
example of a DW theory with matter, now for the particular case when the 
spinor fields are in the adjoint representation of the gauge group. 
The latter theory bears a strong resemblance to the non--abelian 
generalization of the Seiberg--Witten monopol theory. However, twisted $N = 4$ 
SYM does not lead to a topological matter action having a $N_T = 2$ 
supersymmetry. This is due to the fact that $N = 4$ SYM cannot be coupled 
to a $N = 4$ matter hypermultiplet. 

In continuing earlier studies \cite{26} of topological gauge theories
we pursued further the idea of constructing TQFT's with matter leading,
in particular, to a new topological tensor matter action with extended, 
$N_T = 2$, supersymmetry. 

The outline of the paper is as follows: In Sect. 2 we briefly review the
DW theory and then we construct a tensor matter action with $N_T = 1$ 
topological supersymmetry; it is shown that its tensorial structure is 
uniquely fixed by gauge and local Weyl invariance. In Sect. 3 we generalize 
the previous construction for complexified gauge fields, $A_\mu \pm i V_\mu$, 
and pass to the $N_T = 1$ super--BF model with matter. 
In Sect. 4, by a suitable deformation of the super--BF model, in the version 
of \cite{25}, we arrive at the $N_T = 2$ topological B--model with 
matter and underlying complexified supersymmetry $Q \pm i \bar{Q}$. 
In the Appendix it is proven, in accordance with \cite{18,25} 
and contrary to some statement in \cite{27,28}, that the on--shell
conditions in the formulation of the B--model cannot be completely lifted 
by using an appropiate set of auxiliary fields. 
   
Throughout the paper we use the following conventions: Greek letters
$\mu, \nu, \ldots$ denote world indices and lower case latin letters 
$a, b, \ldots$ are flat $SO(4)$ tangent space indices. 

\section{Donaldson--Witten theory coupled to tensor matter fields}

Let us first consider the DW theory whose moduli space is the space of 
anti--selfdual instantons. In order to complete the construction of 
that theory --- which has been described in the Introduction --- we must 
specify its configuration space. It consists of the gauge potential 
$A_\mu$, the Grassmann--odd self--dual tensor, vector and scalar fields 
$\chi_{\mu\nu}$, $\psi_\mu$ and $\eta$, respectively, and the Grassmann--even 
scalar fields $\phi$ and $\bar{\phi}$. For the closure of the topological 
superalgebra it is necessary to introduce the bosonic auxiliary self--dual
tensor field $B_{\mu\nu}$. All the fields are in the adjoint representation,
i.e., taking their values in the Lie algebra $Lie(G)$ of some compact 
(semisimple) gauge group $G$. Throughout this paper we adopt the convention 
to choose the generators $T^i \in Lie(G)$ always anti--Hermitean.

The action of DW theory, with a $N_T = 1$ off--shell equivariantly nilpotent 
topological supersymmetry $Q$, adopting the notation of Ref.~\cite{25}, 
can be cast into the $Q$--exact form
\begin{equation}
\label{2.1}
S_{\rm DW} = Q \Psi_{\rm DW},
\end{equation}
with the gauge fermion (see, also, footnote 3 below)
\begin{equation}
\label{2.2}
\Psi_{\rm DW} = \frac{i}{e^2} \int  d^4x\,\sqrt{g}\, {\rm tr} \Bigr\{
\frac{1}{2} \chi^{\mu\nu} F_{\mu\nu} + 
\frac{i}{4} \chi^{\mu\nu} B_{\mu\nu} - \psi^\mu D_\mu \bar{\phi} + 
\frac{i}{2} \eta [ \bar{\phi}, \phi ] \Bigr\},
\end{equation}
where $F_{\mu\nu} = \partial_{[\mu} A_{\nu]} + [ A_\mu, A_\nu ]$ and
$D_\mu = \partial_\mu + [ A_\mu, ~\cdot~]$ are the field strenght and the
covariant derivative in the adjoint representation, respectively;
$e$ is the usual YM coupling constant. 

In (\ref{2.2}) the gauge fermion has been chosen in a Feynman type gauge, 
thereby the first term enforces the localization into the moduli space and the 
third term ensures that pure gauge degrees of freedom are projected out; the 
remaining terms belong to the non--minimal sector and could be droped 
(getting a Landau type gauge). 

The off--shell equivariantly nilpotent $Q$--transformations take the form 
\begin{alignat}{2}
\label{2.3}
&Q g_{\mu\nu} = 0,
&\qquad
&Q \phi = 0,
\nonumber
\\
&Q A_\mu = \psi_\mu,
&\qquad&
Q \psi_\mu = D_\mu \phi,
\nonumber
\\
&Q \bar{\phi} = \eta,
&\qquad&
Q \eta = [ \bar{\phi}, \phi ],
\nonumber
\\
&Q \chi_{\mu\nu} = B_{\mu\nu},
&\qquad&
Q B_{\mu\nu} = [ \chi_{\mu\nu}, \phi ].
\end{alignat}
Therefore, the topological supercharge $Q$ squares to zero only modulo 
field--dependent gauge transformations,
\begin{equation}
\label{2.4}
Q^2 = \delta_G(\phi),
\end{equation}
which are defined by $\delta_G(\omega) A_\mu = D_\mu \omega$ and 
$\delta_G(\omega) X = [ X, \omega ]$ for all the other fields. Hence, all
the local symmetries of the action, apart from the ordinary gauge invariance,
have been fixed.

Spelling out the action (\ref{2.1}) explicitly one obtains,
recalling that $\chi_{\mu\nu}$ and $B_{\mu\nu}$ are self--dual,
\footnote{The various factors of $i$ are due to some subleties in the 
formulation of topological gauge theories (see, the remarks at the end of
Sect.~3). Formally, they can be avoided completely when the fields 
$\chi_{\mu\nu}$, $\bar{\phi}$, $\eta$ and $B_{\mu\nu}$ in the DW theory and,
later on, $\bar{\eta}$ and $Y$ in the super--BF and the B--model 
(see, Sects.~3 and 4), are redefined by multiplying them with $-i$. Then, 
the ghost number symmetry group changes into $SO(2)$ instead of being 
$SO(1,1)$. The Euclideanized amplitudes are defined, as usual, by 
${\rm exp}(- S)$.}

\begin{align}
\label{2.5}
S_{\rm DW} = \frac{i}{e^2} \int  d^4x\,\sqrt{g}\, {\rm tr} \Bigr\{&
\frac{1}{2} B^{\mu\nu} F_{\mu\nu} - \chi^{\mu\nu} D_\mu \psi_\nu + 
\frac{i}{4} B^{\mu\nu} B_{\mu\nu} -
\frac{i}{4} \phi \{ \chi^{\mu\nu}, \chi_{\mu\nu} \}
\nonumber
\\
& - D^\mu \bar{\phi} D_\mu \phi +
\bar{\phi} \{ \psi^\mu, \psi_\mu \} + \psi^\mu D_\mu \eta + 
\frac{i}{2} [ \bar{\phi}, \phi ]^2 - 
\frac{i}{2} \phi \{ \eta, \eta \} \Bigr\}.
\end{align}
The $Q$--exactness of the action (\ref{2.1}) is common to all Witten type 
topological theories and has striking consequences on the general features
of cohomological gauge theories. It means that the physical observables, in 
particular the partition functions themself, have no dependence on the 
metric $g_{\mu\nu}$ and on the coupling constant $e$.

Now, we describe the inclusion of a new type of interaction into topological 
gauge theories involving antisymmetric tensor matter fields. The way of 
constructing such tensor interactions is governed by the following strategy: 

First, we generalize the coupling (\ref{1.2}) such that it might be 
interpreted as a $\varphi^4$--type theory for antisymmetric tensor matter 
fields leading to the non--abelian extension \cite{29} of the Avdeev--Chizhov 
(AC) model \cite{19} in Minkowski space--time:
\begin{equation}
\label{2.6}
S_{\rm AC}(\alpha) = \int_M d^4x\, \Bigr\{
( D^a \varphi_{ac} ) ( D_b \varphi^{bc} )^\dagger + 
\alpha ( \varphi_{ac} \varphi^{\dagger bc} )
( \varphi^{ad} \varphi_{bd}^\dagger ) \Bigr\}.
\end{equation}
Here, $D^a \varphi_{ab} = \partial^a \varphi_{ab} - \varphi_{ab} A^a$ is the 
covariant derivative of $\varphi_{ab}$ which belongs to some finite (complex)
representation of $Lie(G)$. Remind that the gauge potential $A_a$ is 
anti--Hermitean. For notational simplicity we also droped the group index of 
the matter fields. $\alpha$ is the coupling constant of the quartic 
self--interaction.

The action (\ref{2.6}) is invariant under the following gauge 
transformations, 
\begin{equation*}
\delta_G(\omega) A_\mu = D_\mu \omega,
\qquad
\delta_G(\omega) \varphi_{ab} = \varphi_{ab} \omega,
\qquad
\delta_G(\omega) \varphi_{ab}^\dagger = - \omega \varphi_{ab}^\dagger,
\end{equation*}
where the choice of a complex representation allows for a non--trivial
mixing between the chiral components $(T_{ab}, \tilde{T}_{ab})$ of the
complex tensor field $\varphi_{ab} = T_{ab} + i \tilde{T}_{ab}$
\cite{29}. Let us recall that Minkowski space--time does not allow for 
self--dual fields $\varphi_{ab} \neq \tilde{\varphi}_{ab} \equiv 
(i/2) \epsilon_{abcd} \varphi^{cd}$ due to 
$\Tilde{\Tilde{\varphi}}_{ab} = - \varphi_{ab}$. 

Second, we perform in (\ref{2.6}) a Wick rotation to the Euclidean space 
as a result of which $\varphi_{ab}$ becomes two times the anti--selfdual 
part of $T_{ab}$. Then, after appropriate rescaling of $e$ and $\alpha$, 
we put the resulting action, denoted by $S(\alpha)$, on a general Riemannian 
4--manifold endowed with a vierbein $e_\mu^{\!~~a}$ and a spin connection 
$\omega_\mu^{\!~~ab}$,
\begin{equation}
\label{2.7}
S(\alpha) = \frac{1}{e^2} \int  d^4x\,\sqrt{g}\, \Bigr\{
( \nabla^\mu T_{\mu\rho}^- ) ( \nabla_\nu T^{\nu\rho}_+ ) + \alpha 
( T_{\mu\rho}^- T^{\nu\rho}_+ ) ( T^{\mu\sigma}_- T_{\nu\sigma}^+ ) \Bigr\},
\end{equation}
where $T_{\mu\nu}^\pm$ are the (anti)self--dual parts of the tensor matter 
field $T_{\mu\nu}$,
\begin{equation*}
T_{\mu\nu}^\pm = \frac{1}{2} ( T_{\mu\nu} \pm \tilde{T}_{\mu\nu} ),
\qquad
\tilde{T}_{\mu\nu} = \frac{1}{2} \sqrt{g} \epsilon_{\mu\nu\rho\sigma}
T^{\rho\sigma},
\qquad
\Tilde{\Tilde{T}}_{\mu\nu} = T_{\mu\nu},
\end{equation*}
with the Levi--Civita tensor density being normalized as
\begin{equation*}
\sqrt{g} \epsilon_{\mu\nu\rho\sigma} \epsilon^{abcd} = 
e_{[\mu}^{~~a} \ldots e_{\sigma]}^{~~d},
\qquad
e_\mu^{\!~~a} e_\nu^{\!~~b} g^{\mu\nu} = \delta^{ab},
\qquad 
e_\mu^{\!~~a} e_\nu^{\!~~b} \delta_{ab} = g_{\mu\nu}.
\end{equation*}
As usual, the gauge and metric covariant derivative of $A_\mu$ is defined by
\begin{equation*}
\nabla_\mu = D_\mu + \frac{1}{2} \omega_\mu^{\!~~ab} \sigma_{ab},
\qquad
\omega_\mu^{\!~~ab} = - ( \partial_\mu e_\nu^{\!~~a} - 
\Gamma_{\mu\nu}^\rho e_\rho^{\!~~a} ) e^{\nu b},
\qquad
\partial_{[\mu} e_{\nu]}^{~~a} + \omega_{[\mu}^{~~ab} e_{\nu] b} = 0,
\end{equation*}
with $\sigma_{ab}$ being the generators of the holonomy group $SO(4)$; the 
Levi--Civita connection $\Gamma_{\mu\nu}^\rho$ is determined, as usual, by 
requiring covariant constancy of the metric and absence of torsion. With 
these definitions one gets
\begin{equation*}
\nabla^\mu T_{\mu\nu}^\pm = 
\frac{1}{\sqrt{g}} D^\mu ( \sqrt{g} T_{\mu\nu}^\pm ),
\qquad
D^\mu T_{\mu\nu}^- = \partial^\mu T_{\mu\nu}^- - T_{\mu\nu}^- A^\mu,
\qquad
D^\mu T_{\mu\nu}^+ = \partial^\mu T_{\mu\nu}^+ + A^\mu T_{\mu\nu}^+,
\end{equation*}
where $\Gamma_{\mu\nu}^\nu = \partial_\mu {\rm ln} \sqrt{g}$ has been taken 
into account. Now, it is easy to verify that the action (\ref{2.7}) is 
invariant under the following gauge transformations:
\begin{equation}
\label{2.8}
\delta_G(\omega) A_\mu = D_\mu \omega,
\qquad
\delta_G(\omega) T_{\mu\nu}^- = T_{\mu\nu}^- \omega,
\qquad
\delta_G(\omega) T_{\mu\nu}^+ = - \omega T_{\mu\nu}^+.
\end{equation}

Besides the gauge symmetry, the action (\ref{2.7}) possesses also a discrete
symmetry under Hermitean conjugation, steming from the CP invariance of
the original Avdeev--Chizhov action. Under this conjugation $A_\mu$ and 
$T_{\mu\nu}^\pm$ transform into $- A_\mu$ and $T_{\mu\nu}^\mp$. 

Although, for non--trivial $A_\mu \neq 0$ and without any restriction of 
the holonomy group of the underlying 4--manifold, the coupling in (\ref{2.7}) 
is no more conformally, the action (\ref{2.7}) exhibits still an invariance 
under the following local rescalings of the metric $g_{\mu\nu}$ and the 
tensor matter fields $T_{\mu\nu}^\pm$, 
\begin{equation}
\label{2.9}
\delta_{\rm W}(\sigma) A_\mu = 0,
\qquad
\delta_{\rm W}(\sigma) g_{\mu\nu} = - 2 \sigma g_{\mu\nu},
\qquad
\delta_{\rm W}(\sigma) T_{\mu\nu}^\pm = \sigma T_{\mu\nu}^\pm,
\end{equation}
with $\delta_{\rm W}(\sigma) \sqrt{g} = - ( \sqrt{g}/2 ) g^{\mu\nu}
\delta_{\rm W}(\sigma) g_{\mu\nu} = 4 \sigma \sqrt{g}$. Hence, this action 
satisfies, by construction, one of the important properties of cohomological
gauge theories, namely local scale (or Weyl) invariance.

Third, without spoiling its gauge and local scale invariance, a $N_T = 1$ 
supersymmetric extension of the action (\ref{2.7}) is obtained by introducing
Grassmann--odd (anti)self--dual tensor fields\break $\lambda_{\mu\nu}^\pm$ 
and vector fields $\xi_\nu^\pm$ as the superpartners of 
$T_{\mu\nu}^\pm$ and $\nabla^\mu T_{\mu\nu}^\pm$, respectively, and the 
Grass-mann--even symmetric tensor field 
$\zeta_\mu^{\!\!~~\nu}$ as the superpartner of the gauge invariant expression 
$T_{\mu\rho}^- T^{\nu\rho}_+$. For the closure of the topological 
superalgebra it is necessary to introduce the bosonic auxiliary vector and
symmetric tensor fields $Y_\mu^\pm$ and $G_\mu^{\!\!~~\nu}$,
respectively. This supersymmetric action, denoted by $S_{\rm T}(\alpha)$, 
can be cast, analogous to (\ref{2.1}), in a $Q$--exact form, 
\begin{equation}
\label{2.10}
S_{\rm T}(\alpha) = Q \Psi_{\rm T}(\alpha),
\end{equation}
with the following gauge and locally scale invariant matter fermion
(cf., Eqs.~(19)),
\begin{equation}
\label{2.11}
\Psi_{\rm T}(\alpha) = \frac{1}{2 e^2} \int d^4x\,\sqrt{g}\, \Bigr\{
\xi^\nu_- ( \nabla^\mu T_{\mu\nu}^+ - Y_\nu^+ ) + 
( \nabla_\mu T^{\mu\nu}_- - Y^\nu_- ) \xi_\nu^+ +
\alpha \zeta^\mu_{\!~~\nu} ( 
T_{\mu\rho}^- T^{\nu\rho}_+ - G_\mu^{\!\!~~\nu} ) \Bigr\}.
\end{equation}
The off--shell equivariantly nilpotent $Q$--transformations of the 
matter fields are given by
\begin{alignat}{2}
\label{2.12}
&Q T_{\mu\nu}^- = \lambda_{\mu\nu}^-,
&\qquad
&Q \lambda_{\mu\nu}^- = T_{\mu\nu}^- \phi,
\nonumber
\\
&Q \xi_\nu^- = \nabla^\mu T_{\mu\nu}^- + Y_\nu^-,
&\qquad
&Q Y_\nu^- = \xi_\nu^- \phi - \nabla^\mu \lambda_{\mu\nu}^- +
T_{\mu\nu}^- \psi^\mu,
\nonumber
\\
&Q T_{\mu\nu}^+ = \lambda_{\mu\nu}^+,
&\qquad
&Q \lambda_{\mu\nu}^+ = - \phi T_{\mu\nu}^+,
\nonumber
\\
&Q \xi_\nu^+ = \nabla^\mu T_{\mu\nu}^+ + Y_\nu^+,
&\qquad
&Q Y_\nu^+ = - \phi \xi_\nu^+ - \nabla^\mu \lambda_{\mu\nu}^+ -
\psi^\mu T_{\mu\nu}^+,
\nonumber
\\
&Q \zeta_\mu^{\!\!~~\nu} = T_{\mu\rho}^- T^{\nu\rho}_+ + G_\mu^{\!\!~~\nu},
&\qquad
&Q G_\mu^{\!\!~~\nu} = - \lambda_{\mu\rho}^- T^{\nu\rho}_+ -
T_{\mu\rho}^- \lambda^{\nu\rho}_+,
\end{alignat}
with $Q$ satisfying the topological superalgebra (\ref{2.4}). The 
gauge transformations of $\lambda_{\mu\nu}^\pm,\, \xi_\mu^\pm$ and 
$Y_\mu^\pm$ agree with those of $T_{\mu\nu}^\pm$ (c.f.,~Eqs.~(\ref{2.8})),
and $G_{\mu\nu}$ and $\zeta_{\mu\nu}$ are gauge invariant. 

Performing the $Q$--transformation, thereby making use of Eqs. (\ref{2.3}) 
and (\ref{2.12}), the action (\ref{2.10}) becomes
\begin{align}
\label{2.13}
S_{\rm T}(\alpha) = \frac{1}{e^2} \int d^4x\,\sqrt{g}\, \Bigr\{&
( \nabla^\mu T_{\mu\rho}^- ) ( \nabla_\nu T^{\nu\rho}_+ ) - Y_\mu^- Y^\mu_+ 
\nonumber
\\
& + ( \nabla^\mu \lambda_{\mu\nu}^- - T_{\mu\nu}^- \psi^\mu ) \xi^\nu_+ - 
\xi_\nu^- \phi \xi^\nu_+ -
\xi_\nu^- ( \nabla_\mu \lambda^{\mu\nu}_+ + \psi_\mu T^{\mu\nu}_+ )  
\nonumber
\\
& + \alpha \Bigr(
( T_{\mu\rho}^- T^{\nu\rho}_+ ) ( T^{\mu\sigma}_- T_{\nu\sigma}^+ ) -
\zeta^\mu_{\!\!~~\nu} ( \lambda_{\mu\rho}^- T^{\nu\rho}_+ -
T_{\mu\rho}^- \lambda^{\nu\rho}_+ ) -
\frac{1}{2} G_{\mu\nu} G^{\mu\nu} \Bigr) \Bigr\}.
\end{align}
It should be stressed that, unlike the DW theory, the $Q$--transformations 
(\ref{2.12}) are not obtained from a $N = 2$ supersymmetric tensor matter 
action via a topological twist. Let us shortly comment on why such a 
`detour', if possible at all, was actually not necessary.
This is simply due to the fact that the scalar field $\phi$, entering into 
the DW theory, is $Q$--inert. Therefore, by choosing a suitable gauge and 
locally scale invariant fermion $\Psi_{\rm T}(\alpha)$ and off--shell 
equivariantly nilpotent transformation rules for the matter fields the 
resulting action $S_{\rm T}(\alpha)$ is very alike the action 
$S_{\rm DW}$. However, this is no longer the case for $N_T = 2$ topological 
matter, whose construction is rather involved and, obviously, quite special 
(see, Sect.~4 below). 

The action (\ref{2.6}) of the non-abelian Avdeev--Chizhov model has several 
remarkable features~\cite{29} most of which may be proven also for the action 
(\ref{2.7}). Above all, it is worth notifying that its tensorial structure 
is completely fixed. Namely, due to the following purely algebraic relations,
\begin{equation}
\label{2.14}
g^{\rho\sigma} T_{\mu\rho}^- F^{\mu\nu} T_{\nu\sigma}^+ = 0,
\qquad
T_{\mu\rho}^- ( C^{\mu\rho\nu\sigma} + g^{\rho\sigma} R^{\mu\nu} )
T_{\nu\sigma}^+ = 0
\end{equation}
and
\begin{equation}
\label{2.15}
( T_{\mu\nu}^- T_{\rho\sigma}^+ ) ( T^{\mu\nu}_- T^{\rho\sigma}_+ ) =
4 ( T_{\mu\rho}^- T^{\mu\sigma}_+ ) ( T^{\nu\rho}_- T_{\nu\sigma}^+ ) =
4 ( T_{\mu\rho}^- T^{\mu\sigma}_+ ) ( T_{\nu\sigma}^- T^{\nu\rho}_+ ),
\end{equation}
an {\em unique} gauge and local scale invariant kinetic and quartic 
self--interaction term is singlet out (this explains why the action 
(\ref{2.7}) deserves our interest). In addition, these relations 
forbid the existence of mass and cubic self--interacting terms.

The equivalence of the three possible self--interaction terms (\ref{2.15}) 
can simply be proven by using the identity 
$e^\sigma_{~d} \sqrt{g} \epsilon_{\mu\nu\rho\sigma} \epsilon^{abcd} = 
e_{[\mu}^{~~a} e_\nu^{\!~~b} e_{\rho]}^{~~c}$. In the same way one  
verifies the relation
\begin{equation*}
g^{\rho\sigma} T_{\mu\rho}^- [ \nabla^\mu, \nabla^\nu ] T_{\nu\sigma}^+ = 0,
\end{equation*}
which guarantees the uniqueness of the kinetic term. From this relation 
one derives
\begin{equation}
\label{2.16}
g^{\rho\sigma} T_{\mu\rho}^- F^{\mu\nu} T_{\nu\sigma}^+ = 0,
\qquad
T_{\mu\rho}^- ( R^{\mu\rho\nu\sigma} - g^{\rho\sigma} R^{\mu\nu} )
T_{\nu\sigma}^+ = 0,
\qquad
g^{\mu\nu} g^{\rho\sigma} T_{\mu\rho}^- R T_{\nu\sigma}^+ = 0,
\end{equation}
where $R_{\mu\nu} = g^{\rho\sigma} R_{\mu\rho\nu\sigma}$ and
$R = g^{\mu\nu} R_{\mu\nu}$ are the Ricci tensor and the Ricci scalar,
respectively, $R^\mu_{~\rho\nu\sigma} = 
\partial_{[\nu} \Gamma_{\sigma]\rho}^\mu + 
\Gamma_{\lambda[\nu}^\mu \Gamma_{\sigma]\rho}^\lambda$ being the Riemannian 
curvature tensor. Decomposing $R_{\mu\rho\nu\sigma}$ into its irreducible
parts,
\begin{equation*}
R_{\mu\rho\nu\sigma} = C_{\mu\rho\nu\sigma} -
\frac{1}{6} R ( g_{\mu\nu} g_{\rho\sigma} - g_{\mu\sigma} g_{\nu\rho} ) +
\frac{1}{2} ( g_{\mu\nu} R_{\rho\sigma} - g_{\mu\sigma} R_{\nu\rho} -
g_{\nu\rho} R_{\mu\sigma} + g_{\rho\sigma} R_{\mu\nu} ),
\end{equation*}
where the conformal Weyl tensor $C_{\mu\rho\nu\sigma}$ is completely 
traceless, from (\ref{2.16}) one obtains immediatly the relations 
(\ref{2.14}). The fact, that $C_{\mu\rho\nu\sigma}$ appears in (\ref{2.14}) 
only in combination with $R_{\mu\nu}$ is not surprising if one remembers the 
definition of the Euler number,
\begin{equation*}
\chi = \frac{1}{32 \pi^2} \int  d^4x\,\sqrt{g}\, \Bigr\{
C^{\mu\rho\nu\sigma} C_{\mu\rho\nu\sigma} - 2 R^{\mu\nu} R_{\mu\nu} +
\frac{2}{3} R^2 \Bigr\},
\end{equation*}
and, in addition, takes into account the last of the relations (\ref{2.16}).

Let us remark that the second of Eqs.~(\ref{2.16}) along with our choice 
of the matter fermion (\ref{2.11}) ensures also that a coupling of DW theory 
to tensorial matter does not spoil the local scale invariance. Indeed, it is 
simple to check that the actions (\ref{2.5}) and (\ref{2.13}) are left 
invariant by the following local Weyl symmetry,
\begin{gather}
\delta_{\rm W}(\sigma) g_{\mu\nu} = - 2 \sigma g_{\mu\nu},
\nonumber
\\
\begin{alignat}{3}
&\delta_{\rm W}(\sigma) \bar{\phi} = - 2 \sigma \bar{\phi},
&\qquad
&\delta_{\rm W}(\sigma) T_{\mu\nu}^\pm = \sigma T_{\mu\nu}^\pm,
&\qquad
&\delta_{\rm W}(\sigma) Y_\mu^\pm = - \sigma Y_\mu^\pm,
\\
&\delta_{\rm W}(\sigma) \eta = - 2 \sigma \eta,
&\qquad
&\delta_{\rm W}(\sigma) \lambda_{\mu\nu}^\pm = \sigma \lambda_{\mu\nu}^\pm,
&\qquad
&\delta_{\rm W}(\sigma) \xi_\mu^\pm = - \sigma \xi_\mu^\pm,
\nonumber
\end{alignat}
\end{gather}
where we have only written down the non--trivial scale transformations. Obviously, 
this symmetry commutes with the topological supersymmetry $Q$, i.e., it holds 
$[ \delta_{\rm W}(\sigma), Q ] = 0$.

Since the total action, $S_{\rm DW} + S_{\rm T}(\alpha)$, is $Q$--exact, and 
because $g_{\mu\nu}$ is $Q$--inert, it follows immediately that the metric 
variation of that action is also $Q$--exact. Therefore, the full stress tensor 
$\mathbf{T}_{\mu\nu}(\alpha)$ derived from this action is $Q$--exact, as well, 
which is sufficient to ensure that physical observables have no dependence on 
the metric of the underlying manifold. Since this action is locally scale 
invariant, the trace of the stress tensor becomes, on--shell, the divergence 
of a current,
\begin{equation*}
g^{\mu\nu} \mathbf{T}_{\mu\nu}(\alpha) = \frac{2 g^{\mu\nu}}
{\sqrt{g}} \frac{\delta}{\delta g^{\mu\nu}} \left( S_{\rm DW} +
S_{\rm T}(\alpha) \right) \doteq \partial_\mu j^\mu(\alpha).
\end{equation*}
Moreover, since topological gauge theories should not involve arbitrary
parameters --- at least, as long as they do not enter into the 
$Q$--transformations --- physical observables should also not depend on the 
coupling constant $\alpha$ of the self--interaction term. Indeed, the 
$Q$--exactness of the action (\ref{2.10}) --- which, as already emphasized 
above, has far--reaching consequences --- makes it possible to use 
field--theoretical arguments to conclude that the $\alpha$--dependent term 
in (\ref{2.13}) is irrelevant and can be omitted. Therefore, in the following 
considerations we shall take into account only the kinetic term of the 
tensor matter action. 

\section{$N_T = 1$ super--BF model with tensor matter}

In the previous section we extended the DW theory to a topological model 
with tensor matter. Its moduli space remains to be dominated by instantons. 
But, the evaluation of the partition function in the weak coupling limit, 
which is expected to go still through as in Ref. \cite{10}, will now receive 
contributions to the ratios of determinants of the kinetic operators from the 
even and odd integer spin fields ($A_\mu$, $\chi_{\mu\nu}$, $\psi_\mu$) of the 
gauge multiplet as well as from the even and odd integer spin fields 
($T_{\mu\nu}^\pm$, $\lambda_{\mu\nu}^\pm$, $\xi_\mu^\pm$) of the matter 
multiplet. 

Now, as a preliminary stage, we will generalize the previous construction
for the $N_T = 1$ super--BF model whose moduli space is the space of flat 
complexified gauge fields $A_\mu \pm i V_\mu$. The idea behind of this 
intention is that, in principle, such an generalization allows also for  
introducing an extended, $N_T = 2$, supersymmetry. For that purpose, we have
to assume that, apart from the complexified gauge field, all the 
Grassmann--odd fields are complexified ones, as well. In Sect.~4, when 
turning to the $B$--model, it will become obvious that such an 
extension, roughly speaking, amounts to introduce an extended, $N_T = 2$, 
complexified supersymmetry $Q \pm i \bar{Q}$. Usually, extended 
supersymmetries arise when $N_T = 1$ theories 
are formulated on manifolds with reduced holonomy groups, e.g., DW theory on 
K\"ahler manifolds. Therefore, the complexified supersymmetry 
$Q \pm i \bar{Q}$ which we encounter here is of a different kind as both, 
$Q$ and $\bar{Q}$, must have the same ghost number.

The $N_T = 1$ super--BF model was described in detail in \cite{17} using 
the Batalin--Vilkovsky formalism. However, due to some redundancy in that 
description it is possible to find a more simpler formulation of the model 
with a reduced field content \cite{25}. It consists of the gauge field 
$A_\mu$, the vector field $V_\mu$, the complex Grassmann--odd tensor, vector 
and scalar fields $\chi_{\mu\nu}$, $\psi_\mu$, $\bar{\psi}_\mu$ and 
$\eta$, $\bar{\eta}$, respectively, and the Grassmann--even scalar fields 
$\phi$ and $\bar{\phi}$. Moreover, in order to ensure off--shell equivariantly 
nilpotency of the topological supersymmetry $Q$, $Q^2 = \delta_G(\phi)$, it is
necessary to introduce the bosonic auxiliary tensor and scalar fields
$B_{\mu\nu}$ and $Y$. (Let us point out that $\chi_{\mu\nu}$ and $B_{\mu\nu}$ 
are not self--dual.)

Adopting that formulation, the action of the model can be written in a 
$Q$--exact form, 
\begin{equation}
\label{3.1}
S_{\rm BF} = Q \Psi_{\rm BF},
\end{equation}
where the gauge fermion (again choosen in a Feynman type gauge) is given by
\begin{align}
\label{3.2}
\Psi_{\rm BF} = \frac{i}{e^2} \int d^4x\,\sqrt{g}\, {\rm tr} \Bigr\{&
\frac{1}{2} \tilde{\chi}^{\mu\nu} F_{\mu\nu} + 
\frac{i}{4} \chi^{\mu\nu} B_{\mu\nu} + \chi^{\mu\nu} D_\mu V_\nu
\nonumber
\\
& - \psi^\mu D_\mu \bar{\phi} + \frac{i}{2} \eta [ \bar{\phi}, \phi ] + 
V^\mu D_\mu \bar{\eta} + \frac{i}{2} \bar{\eta} Y \Bigr\},
\end{align}
and the supersymmetry transformations $Q$ are defined as follows:
\begin{alignat}{2}
\label{3.3}
&Q g_{\mu\nu} = 0,
&\qquad
&Q \phi = 0,
\nonumber
\\
&Q A_\mu = \psi_\mu,
&\qquad&
Q \psi_\mu = D_\mu \phi,
\nonumber
\\
&Q V_\mu = \bar{\psi}_\mu,
&\qquad&
Q \bar{\psi}_\mu = [ V_\mu, \phi ],
\nonumber
\\
&Q \bar{\phi} = \eta,
&\qquad&
Q \eta = [ \bar{\phi}, \phi ],
\nonumber
\\
&Q \bar{\eta} = Y,
&\qquad&
Q Y = [ \bar{\eta}, \phi ],
\nonumber
\\
&Q \chi_{\mu\nu} = B_{\mu\nu},
&\qquad&
Q B_{\mu\nu} = [ \chi_{\mu\nu}, \phi ].
\end{alignat}
The reduced configuration space of the model suggests the existence of 
a discrete symmetry under Hermitean conjugation, namely
\footnote{This symmetry is identical with the Hermitean conjugation introduced 
in Ref.~\cite{18} after an appropriate redefinition of the original fields 
through complexified ones (see also remarks at the end of this Section).}
\begin{align}
\label{3.4}
( A_\mu, \psi_\mu, \bar{\psi}_\mu, \phi, V_\mu, \bar{\phi},
\eta, \bar{\eta}, Y, \chi_{\mu\nu}, B_{\mu\nu} ) &\rightarrow
( - A_\mu, i \bar{\psi}_\mu, - i \psi_\mu, - \phi, - V_\mu, \bar{\phi}, 
- i \bar{\eta}, i \eta, Y, i \tilde{\chi}_{\mu\nu}, - B_{\mu\nu} ),
\nonumber
\\
\epsilon_{\mu\nu\rho\sigma} &\rightarrow - \epsilon_{\mu\nu\rho\sigma},
\end{align}
so that one could expect the occurrence of an extended, $N_T = 2$, 
topological supersymmetry $Q \pm i \bar{Q}$. However, carring out in 
(\ref{3.1}) the $Q$--transformation explicitly, which yields
\begin{align}
\label{3.5}
S_{\rm BF} = \frac{i}{e^2} \int d^4x\,\sqrt{g}\, {\rm tr} \Bigr\{&
\frac{1}{2} \tilde{B}^{\mu\nu} F_{\mu\nu} + 
B^{\mu\nu} D_\mu V_\nu + \frac{i}{4} B^{\mu\nu} B_{\mu\nu} - 
\frac{i}{4} \phi \{ \chi^{\mu\nu}, \chi_{\mu\nu} \}
\nonumber
\\
& - D^\mu \bar{\phi} D_\mu \phi + \bar{\phi} \{ \psi^\mu, \psi_\mu \} - 
\tilde{\chi}^{\mu\nu} D_\mu \psi_\nu + \psi^\mu D_\mu \eta
\phantom{\frac{1}{2}}
\nonumber
\\
& - \chi^{\mu\nu} D_\mu \bar{\psi}_\nu + \bar{\psi}^\mu D_\mu \bar{\eta} - 
\chi^{\mu\nu} [ V_\mu, \psi_\nu ] - \psi^\mu [ V_\mu, \bar{\eta} ]  
\nonumber
\\
& - \frac{i}{2} \phi \{ \bar{\eta}, \bar{\eta} \} +
\frac{i}{2} [ \bar{\phi}, \phi ]^2 - \frac{i}{2} \phi \{ \eta, \eta \} + 
V^\mu D_\mu Y + \frac{i}{2} Y^2 \Bigr\},
\end{align}
it is easily seen that this is only partly the case. Therefore, this action
possesses really only a simple, $N_T = 1$, topological supersymmetry.

Let us now turn to the construction of the matter action. For that purpose 
we incorporate into (\ref{2.7}) the vector field $V_\mu$ in such a way
that the invariance under Hermitean conjugation is preserved. This is
simply achieved by replacing in $\nabla^\mu T_{\mu\nu}^\pm$ the gauge 
field $A_\mu$ through $A_\mu \mp i V_\mu$, respectively. The
$\alpha$--independent part of the resulting action, denoted by 
$S_{\rm C}$ (C standing for `complexified'), takes the form 
\begin{equation}
\label{3.6}
S_{\rm C} = \frac{1}{e^2} \int  d^4x\,\sqrt{g}\, \Bigr\{
( \nabla^\mu T_{\mu\rho}^- - i T_{\mu\rho}^- V^\mu )
( \nabla_\nu T^{\nu\rho}_+ - i V_\nu T^{\nu\rho}_+ ) \Bigr\},
\end{equation}
and is, in fact, invariant under Hermitean conjugation 
$( A_\mu, V_\mu, T_{\mu\nu}^\pm ) \rightarrow 
( - A_\mu, - V_\mu, T_{\mu\nu}^\mp )$.

In order to get the $N_T = 1$ supersymmetric extension of (\ref{3.6}) we make 
the ansatz 
\begin{equation}
\label{3.7}
S_{\rm CT} = Q \Psi_{\rm CT},
\end{equation}
and choose the following gauge and locally scale invariant matter fermion
\begin{equation}
\label{3.8}
\Psi_{\rm CT} = \frac{1}{2 e^2} \int d^4x\,\sqrt{g}\, \Bigr\{
\xi^\nu_- ( \nabla^\mu T_{\mu\nu}^+ - i V^\mu T_{\mu\nu}^+ - Y_\nu^+ ) +
( \nabla^\mu T_{\mu\nu}^- - i T_{\mu\nu}^- V^\mu - Y_\nu^- ) \xi^\nu_+
\Bigr\},
\end{equation}
where the $Q$--transformations are given by
\begin{align}
\label{3.9}
&Q T_{\mu\nu}^- = \lambda_{\mu\nu}^-,
\nonumber
\\
&Q \lambda_{\mu\nu}^- = T_{\mu\nu}^- \phi,
\nonumber
\\
&Q \xi_\nu^- = \nabla^\mu T_{\mu\nu}^- - i T_{\mu\nu}^- V^\mu + Y_\nu^-,
\nonumber
\\
&Q Y_\nu^- = \xi_\nu^- \phi - \nabla^\mu \lambda_{\mu\nu}^- +
i \lambda_{\mu\nu}^- V^\mu + T_{\mu\nu}^- ( \psi^\mu + i \bar{\psi}^\mu ),
\nonumber
\\
&Q T_{\mu\nu}^+ = \lambda_{\mu\nu}^+,
\nonumber
\\
&Q \lambda_{\mu\nu}^+ = - \phi T_{\mu\nu}^+,
\nonumber
\\
&Q \xi_\nu^+ = \nabla^\mu T_{\mu\nu}^+ - i V^\mu T_{\mu\nu}^+ + Y_\nu^+,
\nonumber
\\
&Q Y_\nu^+ = - \phi \xi_\nu^+ - \nabla^\mu \lambda_{\mu\nu}^+ +
i V^\mu \lambda_{\mu\nu}^+ - ( \psi^\mu - i \bar{\psi}^\mu ) T_{\mu\nu}^+.
\end{align} 
With that choice the action (\ref{3.7}) takes the form
\begin{align}
\label{3.10}
S_{\rm CT} = \frac{1}{e^2} \int d^4x\,\sqrt{g}\, \Bigr\{&
( \nabla^\mu T_{\mu\rho}^- - i T_{\mu\rho}^- V^\mu )  
( \nabla_\nu T^{\nu\rho}_+ - i V_\nu T^{\nu\rho}_+ ) - Y_\mu^- Y^\mu_+
\nonumber
\\
& + ( \nabla^\mu \lambda_{\mu\nu}^- - i \lambda_{\mu\nu}^- V^\mu ) \xi^\nu_+ - 
\xi_\nu^- ( \nabla_\mu \lambda^{\mu\nu}_+ - i V_\mu \lambda^{\mu\nu}_+ )
\nonumber
\\
& - T_{\mu\nu}^- ( \psi^\mu + i \bar{\psi}^\mu ) \xi^\nu_+ -
\xi_\nu^- \phi \xi^\nu_+ -
\xi_\nu^- ( \psi_\mu - i \bar{\psi}_\mu ) T^{\mu\nu}_+ \Bigr\},
\end{align}
and one can easily verify that, as promised, this action exhibits a discrete
symmetry under Hermitean conjugation, namely
\begin{equation}
\label{3.11}
( A_\mu, \psi_\mu, \bar{\psi}_\mu, \phi, V_\mu,
T_{\mu\nu}^\pm, \lambda_{\mu\nu}^\pm, Y_\mu^\pm, \xi_\mu^\pm ) \rightarrow
( - A_\mu, i \bar{\psi}_\mu, - i \psi_\mu, - \phi, - V_\mu,
T_{\mu\nu}^\mp, \mp \lambda_{\mu\nu}^\mp, Y_\mu^\mp, \mp \xi_\mu^\mp ),
\end{equation}
which is clearly compatible with (\ref{3.4}). Also this symmetry suggests 
the presence of a hidden $N_T = 2$ supersymmetry $Q \pm i \bar{Q}$. However, 
in order to expose such a second supersymmetry $\bar{Q}$ the action 
(\ref{3.5}) must be deformed by adding further terms to the gauge fermion 
(\ref{3.2}) so that its partly discrete symmetry under Hermitean 
conjugation (\ref{3.4}) becomes completely manifest. 
This will be the subject of the next section.  

Finally, there are several points worth to note about the appearence of some
`wrong' signs in the symmetry (\ref{3.4}): 
First, because all of the (real parts of the) fields of the BF--model are 
represented by 
anti--Hermitean matrices, there is an extra minus sign in these 
transformations, e.g., 
$A_\mu \pm i V_\mu \rightarrow - ( A_\mu \mp i V_\mu )$. Second,
in order to ensure that the transformation  
$\chi_{\mu\nu} \pm i \tilde{\chi}_{\mu\nu} \rightarrow
i ( \tilde{\chi}_{\mu\nu} \pm i \chi_{\mu\nu} )$ of the complexified tensor 
fields agrees, formally, with that of the complexified vector fields, 
$\psi_\mu \pm i \bar{\psi}_\mu \rightarrow
i ( \bar{\psi}_\mu \pm i \psi_\mu )$, it is necessary that Hermitean 
conjugation is combined with a simultaneous replacement   
$\epsilon_{\mu\nu\rho\sigma} \rightarrow - \epsilon_{\mu\nu\rho\sigma}$,
which reverses the orientation of the 4--manifold. Third, the transformations 
$\bar{\phi} \rightarrow \bar{\phi}$,
$\eta \pm i \bar{\eta} \rightarrow - i ( \bar{\eta} \pm i \eta )$ and
$Y \rightarrow Y$ have apparently a wrong sign. These extra sign changes 
can be traced back to some subtleties in the formulation of cohomological 
gauge theories. 
The DW theory was originally derived from the Wick--rotated $N = 2$ SYM with 
compact $R$--symmetry group $U(2) \cong Sp(2) \otimes U(1)$, with 
$U(1) \cong SO(2)$ being the ghost number symmetry \cite{10}. But, in 
this approach, the sign of the kinetic term of one 
of the two original scalars, $\phi$ and $\bar{\phi}$, must be changed, so 
that the twisted theory has an $SO(1,1)$ ghost number symmetry \cite{10}. 
One can simply overcome this difficulty if the `problematic' scalar field 
$\phi$ is replaced by $\phi \rightarrow i \phi$. Therefore, under 
Hermitean conjugation the scalars $\phi$ and $\bar{\phi}$ transform like a 
real and a purely imaginary field, respectively. The situation is quite 
similar for the super--BF model. 
Recently, it has been shown that one can completely sidestep this problem 
by twisting directly the Euclidean $N = 2$ SYM with non--compact 
$R$--symmetry group $Sp(2) \otimes SO(1,1)$ instead of 
$Sp(2) \otimes SO(2)$ \cite{30}.

\section{ $N_T = 2$ topological B--model with tensor matter}

The purpose of this Section is to deform the action (\ref{3.5}) of the 
super--BF model --- according to the proposal \cite{25} of Blau and 
Thompson --- to that of the B--model and to reveal the second topological 
supersymmetry $\bar{Q}$ of the matter action (\ref{3.10}). We begin by shortly 
reviewing the structure of the B--model \cite{18}. This model is 
obtained by a certain twist of $N = 4$ SYM, namely breaking down the 
$R$--symmetry group $SU(4)$ to 
$SO(4) \cong Sp(2)_A \otimes Sp(2)_B \otimes U(1)$ and by changing the 
action of the rotation group 
$SO(4) \cong SU(2)_L \otimes SU(2)_R$ of the Euclidean space by replacing 
$SU(2)_L$ and $SU(2)_R$ through the diagonal subgroup of 
$SU(2)_L \otimes Sp(2)_A$ and of $SU(2)_R \otimes Sp(2)_B$, respectively.

The $N = 4$ SYM is believed to be exactly finite and conformal invariant,
even non--perturba-tively \cite{31}. Furthermore, it is believed that the
S--duality \cite{32} in $N = 4$ SYM, which includes a discrete $Z_2$ symmetry 
corresponding to an interchange of electric and magnetic charges along with 
an interchange of weak and strong coupling, $e \rightarrow 1/e$, is exact. 
(It is natural to conjecture that the twisted theory has also an S--duality 
symmetry.) 

For $N = 4$ SYM it is possible to introduce a further coupling constant 
$\theta$ by adding to the YM action a topological term (owing to the absence 
of a chiral anomaly in $N = 4$ SYM it is impossible to shift away the 
$\theta$--term by means of a chiral rotation). 
In the presence of a non--zero $\theta$--angle the original $Z_2$ symmetry 
$e \rightarrow 1/e$ is extended to a full $SL(2,Z)$ symmetry acting on 
$\tau = \theta / 2 \pi + 4 \pi i / e^2$. Thereby, one expects that under
an inversion $\tau \rightarrow - 1/\tau$ of this coupling the gauge group 
$G$ is exchanged with its dual group $Dual(G)$. Moreover, as pointed out 
by 't Hooft \cite{33}, one can consider topological non--trivial gauge 
transformations of $G/Center(G)$ with discrete magnetic 't Hooft flux through 
certain two--cycles of the 4--manifold.

The action of the B--model in the presence of a non--zero $\theta$--term can 
be cast into the form  
\begin{equation}
\label{4.1}
S_{\rm BT}(\theta) = Q \Psi_{\rm BT} - i \theta k =
\bar{Q} \bar{\Psi}_{\rm BT} - i \theta k,
\qquad
k = \frac{1}{32 \pi^2} \int  d^4x\,\sqrt{g}\, {\rm tr} \Bigr\{
\tilde{F}^{\mu\nu} F_{\mu\nu} \Bigr\},
\end{equation}
where $k$ is the instanton number, i.e., its $\theta$--independent part 
is $Q$-- and $\bar{Q}$--exact, but not $Q \bar{Q}$--exact, 
$S_{\rm BT}(0) \neq Q \bar{Q} \Omega_{\rm BT}$ (notice that under the 
Hermitean conjugation, Eqs.~(\ref{3.4}), $k$ transforms into $- k$ whereas 
$\theta$ remains inert). 

The gauge fermion $\Psi_{\rm BT}$ in (\ref{4.1}) is an appropiate extension of
the gauge choice (\ref{3.2}), 
\begin{align}
\label{4.2}
\Psi_{\rm BT} = \frac{i}{e^2} \int  d^4x\,\sqrt{g}\, {\rm tr} \Bigr\{&
\frac{1}{2} \tilde{\chi}^{\mu\nu} ( F_{\mu\nu} - [ V_\mu, V_\nu ] ) + 
\frac{i}{4} \chi^{\mu\nu} ( B_{\mu\nu} - i D_{[\mu} V_{\nu]} ) 
\nonumber
\\
& - \psi^\mu D_\mu \bar{\phi} - \bar{\psi}^\mu [ V_\mu, \bar{\phi} ] + 
\frac{i}{2} \eta [ \bar{\phi}, \phi ] +
V^\mu D_\mu \bar{\eta} + \frac{i}{2} \bar{\eta} Y \Bigr\},
\end{align}
whereas 
\begin{align}
\label{4.3}
\bar{\Psi}_{\rm BT} = \frac{i}{e^2} \int  d^4x\,\sqrt{g}\, {\rm tr} \Bigr\{&
\frac{1}{2} \chi^{\mu\nu} ( F_{\mu\nu} - [ V_\mu, V_\nu ] ) + 
\frac{i}{4} \tilde{\chi}^{\mu\nu} ( B_{\mu\nu} - i D_{[\mu} V_{\nu]} ) 
\nonumber
\\
& - \bar{\psi}^\mu D_\mu \bar{\phi} + \psi^\mu [ V_\mu, \bar{\phi} ] + 
\frac{i}{2} \bar{\eta} [ \bar{\phi}, \phi ] -
V^\mu D_\mu \eta - \frac{i}{2} \eta Y \Bigr\},
\end{align}
is obtained from (\ref{4.2}) by applying the following discrete $Z_2$ symmetry 
of the B--model \cite{25} (see Eq.~(\ref{4.8}) below),
\begin{equation}
\label{4.4}
( A_\mu, \psi_\mu, \bar{\psi}_\mu, \phi, V_\mu, \bar{\phi}, 
\eta, \bar{\eta}, Y, \chi_{\mu\nu}, B_{\mu\nu} )
\rightarrow
( A_\mu, \bar{\psi}_\mu, \psi_\mu, \phi, - V_\mu, \bar{\phi}, 
\bar{\eta}, \eta, - Y, \tilde{\chi}_{\mu\nu}, B_{\mu\nu} ).
\end{equation}

The on--shell equivariantly nilpotent $Q$-- and $\bar{Q}$--transformations,
being interchanged by the $Z_2$ symmetry (\ref{4.4}), are the following:
\begin{alignat}{2}
\label{4.5}
&Q \phi = 0,
&\qquad&
\bar{Q} \phi = 0,
\nonumber
\\
&Q A_\mu = \psi_\mu,
&\qquad&
\bar{Q} A_\mu = \bar{\psi}_\mu,
\nonumber
\\
&Q \psi_\mu = D_\mu \phi,
&\qquad&
\bar{Q} \bar{\psi}_\mu = D_\mu \phi,
\nonumber
\\
&Q V_\mu = \bar{\psi}_\mu,
&\qquad&
\bar{Q} V_\mu = - \psi_\mu,
\nonumber
\\
&Q \bar{\psi}_\mu = [ V_\mu, \phi ],
&\qquad&
\bar{Q} \psi_\mu = - [ V_\mu, \phi ],
\nonumber
\\
&Q \bar{\phi} = \eta,
&\qquad&
\bar{Q} \bar{\phi} = \bar{\eta},
\nonumber
\\
&Q \eta = [ \bar{\phi}, \phi ],
&\qquad&
\bar{Q} \bar{\eta} = [ \bar{\phi}, \phi ],
\nonumber
\\
&Q \bar{\eta} = Y,
&\qquad&
\bar{Q} \eta = - Y,
\nonumber
\\
&Q Y = [ \bar{\eta}, \phi ],
&\qquad&
\bar{Q} Y = - [ \eta, \phi ],
\nonumber
\\
&Q \chi_{\mu\nu} = B_{\mu\nu} + i D_{[\mu} V_{\nu]},
&\qquad&
\bar{Q} \tilde{\chi}_{\mu\nu} = B_{\mu\nu} - i D_{[\mu} V_{\nu]},
\nonumber
\\
&Q B_{\mu\nu} = [ \chi_{\mu\nu}, \phi ] -
i D_{[\mu} \bar{\psi}_{\nu]} - i [ \psi_{[\mu}, V_{\nu]} ],
&\qquad&
\bar{Q} B_{\mu\nu} = [ \tilde{\chi}_{\mu\nu}, \phi ] -
i D_{[\mu} \psi_{\nu]} + i [ \bar{\psi}_{[\mu}, V_{\nu]} ].
\end{alignat}
Here, both operators $Q$ and $\bar{Q}$ square to zero modulo field--dependent 
gauge transformations $\delta_G(\phi)$ but anticommute only {\it on--shell} 
on $\chi_{\mu\nu}$ and $B_{\mu\nu}$,
\begin{equation}
\label{4.6}
Q^2 = \delta_G(\phi),
\qquad
\{ Q, \bar{Q} \} \doteq 0,
\qquad
\bar{Q}^2 = \delta_G(\phi). 
\end{equation}

In the Appendix we reanalyse in detail a statement of Lozano \cite{27}, 
namely that the action of the B--model in the presence of a non--zero 
$\theta$--term can also be cast into the following form:  
\begin{equation}
\label{4.7}
S_{\rm L}(\tau) = Q \Psi_{\rm L} + 2 \pi i \tau k =
\bar{Q} \bar{\Psi}_{\rm L} + 2 \pi i \bar{\tau} k,
\qquad
\tau = \frac{\theta}{2 \pi} + \frac{4 \pi i}{e^2},
\end{equation}
$\bar{\tau}$ being the complex conjugate of $\tau$, with $Q$ and $\bar{Q}$
satisfying the superalgebra (\ref{4.6}) {\it off--shell}. We have not 
been able to confirm that result. A closer analysis (see Appendix) shows 
that, on the one hand, it is always possible to introduce off--shell
equivariantly nilpotent $Q$-- and $\bar{Q}$--transformations, but then 
$S_{\rm L}(\tau)$ cannot be cast into the form (\ref{4.7}).
On the other hand, it is always possible to express $S_{\rm L}(\tau)$
in the form (\ref{4.7}), but then the $Q$-- and $\bar{Q}$--transformations 
close only on--shell. 

In the Feynman type gauge (\ref{4.2}) the $\theta$--independent part of 
the action (\ref{4.1}) reads
\begin{align}
\label{4.8}
S_{\rm BT}(0) = \frac{i}{e^2} \int  d^4x\,\sqrt{g}\, {\rm tr} \Bigr\{&
\frac{1}{2} \tilde{B}^{\mu\nu} ( F_{\mu\nu} - [ V_\mu, V_\nu ] ) + 
\frac{i}{4} D^{[\mu} V^{\nu]} D_{[\mu} V_{\nu]}
\nonumber
\\
& + \frac{i}{4} B^{\mu\nu} B_{\mu\nu} - 
\frac{i}{4} \phi \{ \chi^{\mu\nu}, \chi_{\mu\nu} \} - 
D^\mu \bar{\phi} D_\mu \phi
\nonumber
\\
& + \bar{\phi} \{ \bar{\psi}^\mu, \bar{\psi}_\mu \} - 
[ V^\mu, \bar{\phi} ] [ V_\mu, \phi ] + 
\bar{\phi} \{ \psi^\mu, \psi_\mu \}
\phantom{\frac{1}{2}}
\nonumber
\\
& - \tilde{\chi}^{\mu\nu} D_\mu \psi_\nu +
\tilde{\chi}^{\mu\nu} [ V_\mu, \bar{\psi}_\nu ] + 
\psi^\mu D_\mu \eta - \psi^\mu [ V_\mu, \bar{\eta} ] 
\nonumber
\\
& - \chi^{\mu\nu} D_\mu \bar{\psi}_\nu -
\chi^{\mu\nu} [ V_\mu, \psi_\nu ] + 
\bar{\psi}^\mu D_\mu \bar{\eta} + \bar{\psi}^\mu [ V_\mu, \eta ]
\phantom{\frac{1}{2}}
\nonumber
\\
& - \frac{i}{2} \phi \{ \bar{\eta}, \bar{\eta} \} +
\frac{i}{2} [ \bar{\phi}, \phi ]^2 - \frac{i}{2} \phi \{ \eta, \eta \} + 
V^\mu D_\mu Y + \frac{i}{2} Y^2 \Bigr\}.
\end{align}
By construction, this action is invariant under Hermitean 
conjugation (\ref{3.4}) and may be regarded, formally, as deformation of 
the action (\ref{3.5}) of the $N_T = 1$ super--BF model. By choosing the
Landau type gauge this action and the supersymmetry transformations 
(\ref{4.5}) coincide precisely with those of Ref. \cite{25}.

After having deformed the action of the super--BF model to a point in the 
deformation space where it possesses an extended, $N_T = 2$, supersymmetry, 
let us now show that at this point the matter action (\ref{3.7}) can be cast 
also into a $\bar{Q}$--exact form, 
\begin{equation}
\label{4.9}
S_{\rm CT} = \bar{Q} \bar{\Psi}_{\rm CT},
\end{equation}
with the following choice of the matter fermion,
\begin{equation}
\label{4.10}
\bar{\Psi}_{\rm CT} = - \frac{i}{2 e^2} \int d^4x\,\sqrt{g}\, \Bigr\{
\xi^\nu_- ( \nabla^\mu T_{\mu\nu}^+ - i V^\mu T_{\mu\nu}^+ - Y_\nu^+ ) -
( \nabla^\mu T_{\mu\nu}^- - i T_{\mu\nu}^- V^\mu - Y_\nu^- ) \xi^\nu_+ 
\Bigr\}.
\end{equation}
Indeed, introducing the $\bar{Q}$--transformations of the matter fields 
according to
\begin{align}
\label{4.11}
&\bar{Q} T_{\mu\nu}^- = - i \lambda_{\mu\nu}^-,
\nonumber
\\
&\bar{Q} \lambda_{\mu\nu}^- = i T_{\mu\nu}^- \phi,
\nonumber
\\
&\bar{Q} \xi_\nu^- = i \nabla^\mu T_{\mu\nu}^- + T_{\mu\nu}^- V^\mu + 
i Y_\nu^-,
\nonumber
\\
&\bar{Q} Y_\nu^- = - i \xi_\nu^- \phi + i \nabla^\mu \lambda_{\mu\nu}^- +
\lambda_{\mu\nu}^- V^\mu + T_{\mu\nu}^- ( \bar{\psi}^\mu - i \psi^\mu ),
\nonumber
\\
&\bar{Q} T_{\mu\nu}^+ = i \lambda_{\mu\nu}^+,
\nonumber
\\
&\bar{Q} \lambda_{\mu\nu}^+ = i \phi T_{\mu\nu}^+,
\nonumber
\\
&\bar{Q} \xi_\nu^+ = - i \nabla^\mu T_{\mu\nu}^+ - V^\mu T_{\mu\nu}^+ - 
i Y_\nu^+,
\nonumber
\\
&\bar{Q} Y_\nu^+ = - i \phi \xi_\nu^+ - i \nabla^\mu \lambda_{\mu\nu}^+ -
V^\mu \lambda_{\mu\nu}^+ - ( \bar{\psi}^\mu + i \psi^\mu ) T_{\mu\nu}^+,
\end{align} 
it is easily seen that they are equivariantly nilpotent and anticommute 
with the $Q$--transforma-tions (\ref{3.9}), i.e., $Q$ and $\bar{Q}$ obey the 
topological superalgebra (\ref{4.6}) {\it off--shell}. Then, spelling out 
(\ref{4.9}) in detail, one recovers precisely the orginal action (\ref{3.10}), 
namely
\begin{align}
\label{4.12}
S_{\rm CT} = \frac{1}{e^2} \int d^4x\,\sqrt{g}\, \Bigr\{&
( \nabla^\mu T_{\mu\rho}^- - i T_{\mu\rho}^- V^\mu )  
( \nabla_\nu T^{\nu\rho}_+ - i V_\nu T^{\nu\rho}_+ ) - Y_\mu^- Y^\mu_+
\nonumber
\\
& + ( \nabla^\mu \lambda_{\mu\nu}^- - i \lambda_{\mu\nu}^- V^\mu ) \xi^\nu_+ - 
\xi_\nu^- ( \nabla_\mu \lambda^{\mu\nu}_+ - i V_\mu \lambda^{\mu\nu}_+ )
\nonumber
\\
& - T_{\mu\nu}^- ( \psi^\mu + i \bar{\psi}^\mu ) \xi^\nu_+ -
\xi_\nu^- \phi \xi^\nu_+ -
\xi_\nu^- ( \psi_\mu - i \bar{\psi}_\mu ) T^{\mu\nu}_+ \Bigr\}.
\nonumber
\end{align}

Furthermore, it is simple to verify that the actions (\ref{4.8}) and 
(\ref{3.10}) are invariant under a local rescaling of the metric
$\delta_{\rm W}(\sigma) g_{\mu\nu} = - 2 \sigma g_{\mu\nu}$ and local Weyl
transformations of the fields. The latter are given by
\begin{alignat*}{2}
&\delta_{\rm W}(\sigma) \bar{\phi} = - 2 \sigma \bar{\phi},
&\qquad
&\delta_{\rm W}(\sigma) T_{\mu\nu}^\pm = \sigma T_{\mu\nu}^\pm,
\\
&\delta_{\rm W}(\sigma) \eta = - 2 \sigma \eta,
&\qquad
&\delta_{\rm W}(\sigma) \lambda_{\mu\nu}^\pm = \sigma \lambda_{\mu\nu}^\pm,
\\
&\delta_{\rm W}(\sigma) Y = - 2 \sigma Y,
&\qquad
&\delta_{\rm W}(\sigma) Y_\mu^\pm = - \sigma Y_\mu^\pm,
\\
&\delta_{\rm W}(\sigma) \bar{\eta} = - 2 \sigma \bar{\eta},
&\qquad
&\delta_{\rm W}(\sigma) \xi_\mu^\pm = - \sigma \xi_\mu^\pm,
\end{alignat*}
with the properties $[ \delta_{\rm W}(\sigma), Q ] = 0 
= [ \delta_{\rm W}(\sigma), \bar{Q} ]$, where again we have only written 
down the non--trivial transformations.

This finishes our construction of the topological B--model involving 
antisymmetric tensor matter fields.

In this context, let us mention that recently a fourth, conformal twist of
$N = 4$ SYM has been proposed leading to a conformal invariant deformation 
of the B--model whose action is local scale invariant and has two Weyl 
invariant topological supersymmetries, $Q$ and $\bar{Q}$, \cite{28}.   
Moreover, it has been conjectured that this model could have a dual 
holographic description in the 5--dimensional de Sitter space. We suppose
that it should be possible to couple this model to antisymmetric
tensor matter fields, too.

\section{Concluding remarks}

Motivated by the question whether, at least from a purely algebraic point 
of view, a topological model with matter having $N_T = 2$ supersymmetry can 
be constructed, we have proposed a new type of matter interactions involving 
antisymmetric tensor fields. These interactions may be regarded as 
supersymmetric extensions of a $\varphi^4$--type theory for antisymmetric 
tensor matter fields, firstly considered in \cite{19,29}, on a general, 
curved 4--dimensional Euclidean gravitational background. Such tensorial 
matter interactions have been explicitly worked out for the DW theory, 
the $N_T = 1$ super--BF model and the $N_T = 2$ topological B--model.  

In that paper we have focused primarily on the algebraic aspects of how 
topological gauge theories involving tensor matter fields can be constructed. 
Many other aspects remain still to be clarified. Among the interesting 
questions which deserve further investigations let us only mention the 
following: 
\\
(i) What are the unitarity properties of the independent propagating 
degrees of freedom associated
with antisymmetric tensor matter fields in {\it Euclidean} space?
\\
(ii) What are the relevant equations of the moduli problem in the presence 
of tensor matter fields?
\\
(iii) Are there new topological observables associated with tensor matter 
fields?
\\
(iv) Does, analogous to the DW theory with matter hypermultiplet,
the ghost--number anomaly of the topological B--model 
change when the tensor matter multiplet is coupled?
\\
(v) How behave antisymmetric tensor matter fields in topological gauge 
theories under renormalization? 
\\
(vi) Is it possible to couple tensorial matter also to the $N_T = 2$ 
topological A--model 
whose underlying supersymmetries, $Q$ and $\bar{Q}$, have different ghost 
numbers and therefore different cohomologies?
\\
(vii) Is it possible to construct a $N = 2$ --- or even a $N = 4$ ---
supersymmetric extension of the (non--abelian) Avdeev--Chizhov model from 
which tensor matter interactions, like the ones introduced in this paper, 
could be obtained via a topological twist? 
\\
(viii) Recently, Berhadsky, Sadov and Vafa \cite{16} have shown that all the
topologically twisted $N = 4$ gauge theories appear quite naturally as 
world--volume theories of Dirichlet $p$--brane instantons in string theory. 
With respect to this the perhaps most interesting question is whether 
topological tensor matter interactions could also appear in some low--energy 
effective tensor gauge theories derived from string models.

\begin{appendix}
\section{Lozano's formulation of the B--model}

In this Appendix it will be shown that in Lozano's action (\ref{4.7})
(see, Ref.~\cite{27}, Eq.~(7.17)),
\begin{equation*}
\tag{A.1}
S_{\rm L}(\tau) = Q \Psi_{\rm L} + 2 \pi i \tau k = 
\bar{Q} \bar{\Psi}_{\rm L} + 2 \pi i \bar{\tau} k \equiv 
S_{\rm L}(\theta = 0) + i \theta k,
\end{equation*}
the topological supercharges $Q$ and $\bar{Q}$ do not provide an 
off--shell formulation of the B--model.
First of all, to be in line with the convention used here, let us
perform the following redefinitions of Lozano's fields (denoted
by a subscript L): 
\begin{alignat*}{3}
&[ Q ]_{\rm L} = Q,
&\qquad
&[ A_\mu ]_{\rm L} = - i A_\mu,
&\qquad
&[ B ]_{\rm L} = - i \sqrt{2} \bar{\phi},
\\
&[ \bar{Q} ]_{\rm L} = \bar{Q},
&\qquad
&[ V_\mu ]_{\rm L} = \frac{i}{\sqrt{2}} V_\mu,
&\qquad
&[ C ]_{\rm L} = - i \sqrt{2} \phi,
\\
&[ \chi_{\mu\nu}^\pm ]_{\rm L} = \frac{i}{2} \chi_{\mu\nu}^\pm, 
&\qquad
&[ \psi_\mu ]_{\rm L} = - i \psi_\mu,
&\qquad
&[ \eta ]_{\rm L} = - 2 i \eta,
\\
&[ P ]_{\rm L} = - 4 i Y,
&\qquad
&[ \bar{\psi}_\mu ]_{\rm L} = - i \bar{\psi}_\mu,
&\qquad
&[ \bar{\eta} ]_{\rm L} = - 2 i \bar{\eta},
\qquad
[ N_{\mu\nu}^\pm ]_{\rm L} = i B_{\mu\nu}^\pm,
\end{alignat*}
where the overall factor of $i$ is due to the different choice of the 
group generators, $[ T^i ]_{\rm L} = - i T^i$.

After carrying out these redefinitions in Eqs.~(7.11) and (7.16) of 
Ref.~\cite{27} and coupling the resulting action 
$S_{\rm L}(\theta = 0) + i \theta k$ to Euclidean gravity, which requires a 
non--minimal $R_{\mu\nu}$--dependent term $R^{\mu\nu} V_\mu V_\nu$, for the 
$\theta$--independent part one obtains
\begin{align*}
S_{\rm L}(\theta = 0) = \frac{1}{e^2} \int d^4x\,&\sqrt{g}\, {\rm tr} \Bigr\{
\frac{1}{4} ( F^{\mu\nu} - [ V^\mu, V^\nu ] )
( F_{\mu\nu} - [ V_\mu, V_\nu ] ) + 
\frac{1}{4} D^{[\mu} V^{\nu]} D_{[\mu} V_{\nu]} 
\\
& - 2 \chi^{\mu\nu}_+ D_\mu \psi_\nu +
2 \chi^{\mu\nu}_+ [ V_\mu, \bar{\psi}_\nu ] -
2 \psi^\mu D_\mu \eta - 2 \bar{\psi}^\mu [ V_\mu, \eta ]
\\
& + \frac{1}{2} \phi \{ \chi^{\mu\nu}_+, \chi_{\mu\nu}^+ \} -
2 \bar{\phi} \{ \psi^\mu, \psi_\mu \} + 
2 \phi \{ \eta, \eta \} - \frac{1}{2} B^{\mu\nu}_+ B_{\mu\nu}^+ 
\\
& - 2 \chi^{\mu\nu}_- D_\mu \bar{\psi}_\nu -
2 \chi^{\mu\nu}_- [ V_\mu, \psi_\nu ] -
2 \bar{\psi}^\mu D_\mu \bar{\eta} + 2 \psi^\mu [ V_\mu, \bar{\eta} ]
\\
& + \frac{1}{2} \phi \{ \chi^{\mu\nu}_-, \chi_{\mu\nu}^- \} -
2 \bar{\phi} \{ \bar{\psi}^\mu, \bar{\psi}_\mu \} + 
2 \phi \{ \bar{\eta}, \bar{\eta} \} - \frac{1}{2} B^{\mu\nu}_- B_{\mu\nu}^-
\\
& + 2 D^\mu \bar{\phi} D_\mu \phi + 
2 [ V^\mu, \bar{\phi} ] [ V_\mu, \phi ] - 2 [ \bar{\phi}, \phi ]^2 -
2 V^\mu D_\mu Y - 2 Y^2 \Bigr\},
\tag{A.2}
\end{align*}
where the $R_{\mu\nu}$--dependence is cancelled by the $Y$--dependent terms.
 
Obviously, the action $S_{\rm L}(\theta = 0) + i \theta k$ is 
invariant under a discrete $Z_2$ symmetry acting on both the fields 
and the coupling constants \cite{27}, 
\begin{align*}
( A_\mu, \psi_\mu, \bar{\psi}_\mu, \phi, V_\mu, \bar{\phi}, 
\eta, \bar{\eta}, Y, \chi_{\mu\nu}^\pm, B_{\mu\nu}^\pm )
&\rightarrow
( A_\mu, - \bar{\psi}_\mu, - \psi_\mu, \phi, - V_\mu, \bar{\phi}, 
- \bar{\eta}, - \eta, - Y, - \chi_{\mu\nu}^\mp, - B_{\mu\nu}^\mp )
\\
\tag{A.3}
( \epsilon_{\mu\nu\rho\sigma}, \theta ) 
&\rightarrow
( - \epsilon_{\mu\nu\rho\sigma}, - \theta ).
\end{align*}
Now, decomposing $F_{\mu\nu}$ into its self--dual and anti--selfdual parts,
then the action $S_{\rm L}(\theta = 0)$ can be expressed either as a 
$Q$--exact or as a $\bar{Q}$--exact term, in both cases modulo a term 
depending only on the instanton number $k$, 
\begin{equation*}
\tag{A.4}
S_{\rm L}(\theta =0) = Q \Psi_{\rm L} - \frac{8 \pi^2 k}{e^2} =
\bar{Q} \bar{\Psi}_{\rm L} + \frac{8 \pi^2 k}{e^2}.           
\end{equation*}
Here the gauge fermions are given by
\begin{align*}
\Psi_{\rm L} = \frac{1}{e^2} \int  d^4x\,\sqrt{g}\, {\rm tr} \Bigr\{&
\frac{1}{2} \chi^{\mu\nu}_+ ( F_{\mu\nu}^+ - [ V_\mu, V_\nu ]^+ -
B_{\mu\nu}^+ ) + 2 \psi^\mu D_\mu \bar{\phi} + 
2 \bar{\psi}^\mu [ V_\mu, \bar{\phi} ]
\\
& + \frac{1}{2} \chi^{\mu\nu}_- \bigr( 
( D_{[\mu} V_{\nu]})^- - B_{\mu\nu}^- \bigr) - 2 \eta [ \bar{\phi}, \phi ] + 
V^\mu D_\mu \bar{\eta} - 2 \bar{\eta} Y \Bigr\},
\\
\bar{\Psi}_{\rm L} = \frac{1}{e^2} \int  d^4x\,\sqrt{g}\, {\rm tr} \Bigr\{&
\frac{1}{2} \chi^{\mu\nu}_- ( F_{\mu\nu}^- -
[ V_\mu, V_\nu ]^- + B_{\mu\nu}^- ) + 2 \bar{\psi}^\mu D_\mu \bar{\phi} - 
2 \psi^\mu [ V_\mu, \bar{\phi} ]
\\
& - \frac{1}{2} \chi^{\mu\nu}_+ \bigr( 
( D_{[\mu} V_{\nu]} )^+ - B_{\mu\nu}^+ \bigr) -
2 \bar{\eta} [ \bar{\phi}, \phi ] -
V^\mu D_\mu \eta + 2 \eta Y \Bigr\}.
\end{align*}
Then, recasting within the full action $S_{\rm L}(\theta = 0) + i \theta k$ 
the topological part in terms of the modular coupling
$\tau = \theta / 2 \pi + 4 \pi i / e^2$ one gets directly the action (A.1) 
with a topological term depending only upon $\tau$.

The $Q$-- and $\bar{Q}$--transformations, being interchanged by the $Z_2$
symmetry (A.3), $Q \leftrightarrow - \bar{Q}$, are given by 
\begin{alignat*}{2}
&Q \phi = 0,
&\qquad&
\bar{Q} \phi = 0,
\\
&Q A_\mu = \psi_\mu,
&\qquad&
\bar{Q} A_\mu = \bar{\psi}_\mu,
\\
&Q \psi_\mu = D_\mu \phi,
&\qquad&
\bar{Q} \bar{\psi}_\mu = D_\mu \phi,
\\
&Q V_\mu = \bar{\psi}_\mu,
&\qquad&
\bar{Q} V_\mu = - \psi_\mu,
\\
&Q \bar{\psi}_\mu = [ V_\mu, \phi ],
&\qquad&
\bar{Q} \psi_\mu = - [ V_\mu, \phi ],
\\
&Q \bar{\phi} = \eta,
&\qquad&
\bar{Q} \bar{\phi} = \bar{\eta},
\\
&Q \eta = [ \bar{\phi}, \phi ],
&\qquad&
\bar{Q} \bar{\eta} = [ \bar{\phi}, \phi ],
\\
&Q \bar{\eta} = Y,
&\qquad&
\bar{Q} \eta = - Y,
\\
&Q Y = [ \bar{\eta}, \phi ],
&\qquad&
\bar{Q} Y = - [ \eta, \phi ],
\\
&Q \chi_{\mu\nu}^+ = F_{\mu\nu}^+ - [ V_\mu, V_\nu ]^+ + B_{\mu\nu}^+,
&\qquad&
\bar{Q} \chi_{\mu\nu}^- = F_{\mu\nu}^- - [ V_\mu, V_\nu ]^- - B_{\mu\nu}^-,
\\
&Q B_{\mu\nu}^+ = [ \chi_{\mu\nu}^+, \phi ] -
( D_{[\mu} \psi_{\nu]} - [ \bar{\psi}_{[\mu}, V_{\nu]} ] )^+,
&\qquad&
\bar{Q} B_{\mu\nu}^- = - [ \chi_{\mu\nu}^-, \phi ] +
( D_{[\mu} \bar{\psi}_{\nu]} + [ \psi_{[\mu}, V_{\nu]} ] )^-,
\\
&Q \chi_{\mu\nu}^- = ( D_{[\mu} V_{\nu]} )^- + B_{\mu\nu}^-,
&\quad&
\bar{Q} \chi_{\mu\nu}^+ = - ( D_{[\mu} V_{\nu]} )^+ - B_{\mu\nu}^+,
\\
&Q B_{\mu\nu}^- = [ \chi_{\mu\nu}^-, \phi ] -
( D_{[\mu} \bar{\psi}_{\nu]} + [ \psi_{[\mu}, V_{\nu]} ] )^-,
&\qquad&
\bar{Q} B_{\mu\nu}^+ = - [ \chi_{\mu\nu}^+, \phi ] +
( D_{[\mu} \psi_{\nu]} - [ \bar{\psi}_{[\mu}, V_{\nu]} ] )^+. 
\tag{A.5}
\end{alignat*}
Here, the operators $Q$ and $\bar{Q}$ are both equivariantly nilpotent and 
anticommute on--shell for $\chi_{\mu\nu}^\pm$ and $B_{\mu\nu}^\pm$ and 
off--shell for all the other fields, i.e., unlike to the claim of 
Ref. \cite{27}, they do {\em not} provide an off--shell realization of the 
topological superalgebra,
\begin{equation*}
\tag{A.6}
Q^2 = \delta_G(\phi),
\qquad
\{ Q, \bar{Q} \} \doteq 0,
\qquad
\bar{Q}^2 = \delta_G(\phi). 
\end{equation*}

Here, some remarks are in order: First, in the case under consideration
it is impossible to cast $S_{\rm L}(\theta = 0)$ into the $Q$-- and 
$\bar{Q}$--exact form $Q \bar{Q} \Omega_{\rm L}$ for some gauge boson 
$\Omega_{\rm L}$. Second, whenever $Q$ and $\bar{Q}$ are interchanged by a 
$Z_2$ symmetry the action $S_{\rm L}(\theta = 0)$ can be cast into the form 
(A.4). In the present case this is only possible when $Q$ and $\bar{Q}$ 
anticommute on--shell. Third, it is impossible to replace in (A.4) $Q$ and 
$\bar{Q}$ by off--shell equivariantly nilpotent operators because such a 
replacement would come into conflict with the $Z_2$ symmetry (A.3). Indeed, 
if we replace in (A.5) the $\bar{Q}$--transformations for $\chi_{\mu\nu}^\pm$ 
and $B_{\mu\nu}^\pm$ according to
\begin{align*}
&\bar{Q} \chi_{\mu\nu}^- = F_{\mu\nu}^- - [ V_\mu, V_\nu ]^- - i B_{\mu\nu}^-,
\nonumber
\\
&\bar{Q} B_{\mu\nu}^- = i [ \chi_{\mu\nu}^-, \phi ] -
i ( D_{[\mu} \bar{\psi}_{\nu]} + [ \psi_{[\mu}, V_{\nu]} ] )^-,
\nonumber
\\
&\bar{Q} \chi_{\mu\nu}^+ = - ( D_{[\mu} V_{\nu]} )^+ - i B_{\mu\nu}^+,
\nonumber
\\
&\bar{Q} B_{\mu\nu}^+ = i [ \chi_{\mu\nu}^+, \phi ] -
i ( D_{[\mu} \psi_{\nu]} - [ \bar{\psi}_{[\mu}, V_{\nu]} ] )^+, 
\end{align*}
leaving all the other transformations unaltered, it is easily
seen that the modified $Q$-- and $\bar{Q}$--transformations provide an
off--shell realization of the superalgebra (A.6). However, the old $Q$-- 
and the new $\bar{Q}$--transformations are no longer related to each 
other through a $Z_2$ symmetry and, therefore, the modified 
$\bar{Q}$--transformations are not any more a symmetry of 
$S_{\rm L}(\theta = 0)$!

Finally, to close our analysis, let us state the relation between the action 
and the supersymmetry transformations given in Sect.~4 and those presented 
above. After integrating out in (\ref{4.8}) and (A.2) the auxiliary field 
$B_{\mu\nu}$ and redefining the Blau--Thompson's fields 
(denoted by a subscript BT) according to:  
\begin{alignat*}{3}
&[ Q ]_{\rm BT} = \zeta ( Q - \bar{Q} ),
&\qquad
&[ A_\mu ]_{\rm BT} = A_\mu,
&\qquad
&[ \phi ]_{\rm BT} = - i \phi,
\\
&[ \bar{Q} ]_{\rm BT} = \zeta ( Q + \bar{Q} ),
&\qquad
&[ V_\mu ]_{\rm BT} = V_\mu,
&\qquad
&[ \bar{\phi} ]_{\rm BT} = 2 \bar{\phi},
\\
&[ \chi_{\mu\nu} ]_{\rm BT} = - 2 \zeta \chi_{\mu\nu}, 
&\qquad
&[ \psi_\mu ]_{\rm BT} = \zeta ( \psi_\mu - \bar{\psi}_\mu ),
&\qquad
&[ \eta ]_{\rm BT} = 2 \zeta ( \eta - \bar{\eta} ),
\\
&[ Y ]_{\rm BT} = - 2 i Y,
&\qquad
&[ \bar{\psi}_\mu ]_{\rm BT} = \zeta ( \psi_\mu + \bar{\psi}_\mu ),
&\qquad
&[ \bar{\eta} ]_{\rm BT} = 2 \zeta ( \eta + \bar{\eta} ),
\end{alignat*}
with the abbreviation $\zeta = (1 - i)/2$, one finds that
$S_{\rm BT}(\theta = 0) = - S_{\rm L}(\theta = 0)$. Furthermore, one verifies 
that the transformations (\ref{4.5}) match precisely those given in (A.5).
Hence, the formulations of the B--model proposed by Blau and Thompson 
\cite{25} and by Lozano \cite{27} are equivalent on--shell, but they differ 
from each other after introducing of $B_{\mu\nu}$. But, in neither cases
the on--shell condition (A.6) can be completely lifted by introducing 
off--shell formulations for the topological supersymmetries $Q$ and $\bar{Q}$
using auxiliary fields.
\end{appendix}


\end{document}